\documentclass[a4 paper, 12 pt] {article}
\def\RR{{\rm I\kern-.1567em R}}                              % Doppel R
 \def\CC{{\rm C\kern-4.7pt                                    % Doppel C
 \vrule height 7.7pt width 0.4pt depth -0.5pt \phantom {.}}}
 \def\ZZ{{\sf Z\kern-4.5pt Z}}

\begin{document}

\title{\large{\bf New Integrable Sectors in Skyrme and 4-dimensional $CP^n$
Model}}

\author{\normalsize{C. Adam}$^{a)}$\thanks{adam@fpaxp1.usc.es}, \,  J.
S\'{a}nchez-Guill\'{e}n$^{a)}$\thanks{joaquin@fpaxp1.usc.es} \,
and A.
Wereszczy\'{n}ski$^{b)}$\thanks{wereszczynski@th.if.uj.edu.pl}
      \\ \\
       \small{ $^{a)}$Departamento de Fisica de Particulas, Universidad
       de Santiago}
       \\
       \small{ and Instituto Galego de Fisica de Altas Enerxias (IGFAE)}
       \\ \small{E-15782 Santiago de Compostela, Spain}
       \\ \\ \small{ $^{b)}$Institute of Physics,  Jagiellonian University,}
       \\ \small{ Reymonta 4, 30-059 Krak\'{o}w, Poland} }
\maketitle
\begin{abstract}
The application of a weak integrability concept to the Skyrme and
$CP^n$ models in 4 dimensions is investigated. A new integrable
subsystem of the Skyrme model, allowing also for non-holomorphic
solutions, is derived. This procedure can be applied to the
massive Skyrme model, as well. Moreover, an example of a family of
chiral Lagrangians providing exact, finite energy Skyrme-like
solitons with arbitrary value of the topological charge, is given.
In the case of $CP^n$ models a tower of integrable subsystems is
obtained. In particular, in (2+1) dimensions a one-to-one
correspondence between the standard integrable submodel and the
BPS sector is proved. Additionally, it is shown that weak
integrable submodels allow also for non-BPS solutions. Geometric
as well as algebraic interpretations of the integrability
conditions are also given.
\end{abstract}
\newpage
%%%%%%%%%%%%%%%%%%%%%%%%%%%%%%%%%%%%%%%%%%%%%%%%%%%%%%%%%%%%%%
\section{Introduction}
%%%%%%%%%%%%%%%%%%%%%%%%%%%%%%%%%%%%%%%%%%%%%%%%%%%%%%%%%%%%%%
Among  nonlinear field theories,  many models with topological
solitons appear to possess particular importance in various
branches of physics (see for example \cite{cond mat1}, \cite{cond
mat2}, \cite{other}). In (1+1) dimensions one can take advantage of
the well-known methods to study the mathematical structure and
dynamics of such objects (called kinks or alternatively domain
walls), like
 the inverse scattering method, Lax pair formalism or the
 B\"{a}cklund transformation \cite{soliton methods1},
\cite{soliton methods2}. In
principle, these methods are connected with the integrability
property of models, which implies the existence of  infinitely many
conserved currents. As a consequence, the chances for the
construction of exact, topological solutions are strongly
correlated with the integrability. Unfortunately in higher spatial
dimensions there are no such general tools allowing for a
systematical construction of solitons. What one has instead are
integrable subsectors, like the holomorphic solitons of Belavin
and Polyakov in 3 dimensions, predecessors of the selfdual
instantons of Yang-Mills in 4 dimensions.

However, there is a new promising approach, based on the
construction of local zero curvature representations for nonlinear
models, which enables one to calculate sufficient conditions for
the existence of integrable subsystems, even though the full model
does not have neccessarily such a property (which would be thereby
exhibited if present \cite{integrability1},
\cite{integrability2}). Here, integrability is understood as
possessing  a zero curvature formulation and the existence of an
infinite number of conserved currents. The significance of this
method originates in the expectation that also in higher
dimensional theories integrability is an important step on the way
of deriving analytical solutions.
\\
That this works has been shown in investigations of especially
useful models with Hopf solitons, where an integrable sector of
the Nicole model has been found \cite{nicole}. It was also proved
that the simplest Hopf soliton belongs to this sector. Further, a
model with infinitely many exact hopfions has been given
\cite{aratyn1}, \cite{aratyn2}.

As is well known, a general sufficient condition (so called
generalized BPS) for the Skyrme model
\cite{skyrme1}-\cite{skyrme3} analogous to the Baby Skyrme or
self-dual cases, has not been found. Simplified models based on
the Skyrme idea of avoiding the scaling instabilities, like the
ones mentioned above, have been investigated. Among them, the
interesting Skyrme-Faddeev-Niemi model. Unluckily the integrable
sector of this well-known model \cite{niemi1}, \cite{niemi2} seems to
contain no topological solitons. In order to cope with this
problem, one natural option is to relax the pertinent
integrability condition (which in this case is the eikonal
equation \cite{eikonal1}, \cite{eikonal2}). The construction of such
weaker integrability conditions has been recently described in
\cite{weak}. As one might anticipate, the space of solutions of
this new, weaker integrable submodel is considerably larger than
the standard one. However, the problem whether it is populated by
hopfions is still an open question.
\\
In the case of the Skyrme model the zero curvature method also
allowed to discover a special integrable subsystem
\cite{ferreira1}. The skyrmion with unit topological charge is a
solution of this restricted submodel, but solitons with higher
value of the baryon number do not belong to the integrable
submodel.
\\
The main aim of the present paper is to define, by means of weaker
integrability conditions, a new ('weaker') integrable sector of
the Skyrme theory with a substantially richer  set of solutions.
It is carried out in Section 2, where, in addition, we generalize
the procedure of Aratyn, Ferreira and Zimerman and introduce a
Skyrme-like model which possesses infinitely many exact, chiral
solitons with arbitrary topological charge. Using this toy-model
one can study in an analytical way chiral solitons,
their energies and profiles.
It is an alternative approach to the so-called
Baby Skyrme models \cite{baby1}, \cite{baby2} which also allows
for analytical treatment of the solitons (baby skyrmions) but in
the lower dimensional space-time.
\\
In Section 3, we present the investigations of the $CP^n$ model in
any dimension with similar results. In particular, in the case of
the $CP^n$ model in two spatial dimensions, we discuss connections
between strong/weak integrability and BPS/non-BPS sectors. The
conclusions are summarized in Section 4.
%%%%%%%%%%%%%%%%%%%%%%%%%%%%%%%%%%%%%%%%%%%%%%%%%%%%%%%%%%%%%%
\section{Skyrme model}
%%%%%%%%%%%%%%%%%%%%%%%%%%%%%%%%%%%%%%%%%%%%%%%%%%%%%%%%%%%%%%
%%%%%%%%%%%%%%%%%%%%%%%%%%%%%%%%%%%%%%%%%%%%%%%%
\subsection{Standard integrable submodel}
%%%%%%%%%%%%%%%%%%%%%%%%%%%%%%%%%%%%%%%%%%%%%%%%
The Skyrme model, without potential term for the chiral field, is
given by the following formula
\begin{equation}
L=\frac{f^2_{\pi}}{4} Tr \left( U^{\dagger}
\partial_{\mu} U U^{\dagger} \partial^{\mu} U \right)
-\frac{1}{32e^2} Tr \left[ U^{\dagger} \partial_{\mu} U,
U^{\dagger} \partial_{\mu} U \right]^2, \label{skyrme}
\end{equation}
where $f_{\pi}$, $e$ are constants and $U$ is a $SU(2)$-valued
matrix field parameterized in the standard manner as
\begin{equation}
U=e^{i\xi_i \tau^i}. \label{def U}
\end{equation}
Here, $\tau_i, \; i=1,2,3$ are the Pauli matrices and $\xi_i$ are
real fields. However, using results of Reference \cite{ferreira1}
it is convenient to take advantage of the slightly different
parametrization
\begin{equation}
U=e^{i\xi T}. \label{def U1}
\end{equation}
Here
\begin{equation}
\xi \equiv \sqrt{\xi^2_1+\xi^2_2+\xi^2_3} \label{def xi}
\end{equation}
and
\begin{equation}
T \equiv \frac{1}{1+|u|^2} \left(
\begin{array}{cc}
|u|^2-1&-2iu \\
2iu^*& 1-|u|^2
\end{array}
\right), \label{def T}
\end{equation}
where the complex field appears due to the standard stereographic
projection
\begin{equation}
\frac{\vec{\xi}}{\xi}= \frac{1}{1+|u|^2} \left( u+u^*, -i(u-u^*),
|u|^2-1 \right). \label{stereo}
\end{equation}
Then, in the notation of  \cite{ferreira1} we derive the equations
of motion
\begin{equation}
D^{\mu} B_{\mu} = \partial^{\mu} B_{\mu} + [A^{\mu},B_{\mu}]=0
\label{eq mot}
\end{equation}
where
\begin{equation}
A_{\mu} \equiv \frac{1}{1+|u|^2} \left( -i\partial_{\mu} u \tau_+
- i\partial_{\mu} u^* \tau_- + \frac{1}{2} (u \partial_{\mu} u^* -
u^* \partial_{\mu} u) \tau_3 \right) \label{def A}
\end{equation}
\begin{equation}
B_{\mu} \equiv -i R_{\mu} \tau_3 +\frac{2 \sin \xi}{1+|u|^2}
\left( e^{i\xi} S_{\mu} \tau_+-e^{-i\xi}S^*_{\mu} \tau_- \right),
\label{def B}
\end{equation}
whereas
\begin{equation}
R_{\mu} \equiv \partial_{\mu} \xi -8\lambda \frac{\sin^2
\xi}{(1+|u|^2)^2} (N_{\mu} + N_{\mu}^*) \label{def R}
\end{equation}
\begin{equation}
S_{\mu} \equiv \partial_{\mu} u + 4\lambda \left( M_{\mu}
-\frac{2\sin^2 \xi}{(1+|u|^2)^2} K_{\mu} \right) \label{def S}
\end{equation}
and
\begin{equation}
K_{\mu} \equiv (\partial^{\nu} u \partial_{\nu} u^*)
\partial_{\mu} u -(\partial_{\nu}u)^2 \partial_{\mu} u^* \label{def
K}
\end{equation}
\begin{equation}
M_{\mu} \equiv (\partial^{\nu} \partial_{\nu} \xi) \partial_{\mu}
\xi - (\partial_{\nu}u)^2 \partial_{\mu} u \label{def M}
\end{equation}
\begin{equation}
N_{\mu} \equiv (\partial^{\nu} u \partial_{\nu} u^*)
\partial_{\mu} \xi -(\partial_{\nu} \xi \partial^{\nu} u) \partial_{\mu}
u^* . \label{def N}
\end{equation}
In addition, $\tau_{\pm} \equiv (\tau_1 \pm i \tau_2)/2$.
\\
It has been proved \cite{ferreira1} that the standard integrable
sector of the Skyrme model is defined by imposing two constraints
\begin{equation}
S_{\mu}\partial^{\mu} u =0, \; \; \; R_{\mu} \partial^{\mu} u =0
\label{old int cond1}
\end{equation}
or in a more restricted form
\begin{equation}
(\partial_{\mu}u)^2=0, \;\;\; \partial^{\mu} \xi \partial_{\mu} u
=0 .\label{old int cond2}
\end{equation}
One can notice that the first integrability condition is nothing
else but the complex eikonal equation.
\\
Then, one can construct two classes of infinitely many currents
\begin{equation}
J_{\mu}^G=\frac{\partial G}{\partial u} S_{\mu} - \frac{\partial
G}{\partial u^*} S^*_{\mu} \label{current1}
\end{equation}
\begin{equation}
J_{\mu}^{(H_1,H_2)}=4\sin \xi \cos \xi (H_1S_{\mu}+H_2S_{\mu}^*)
-(1+|u|^2)^2 \left(\frac{\partial H_1}{\partial u^*} +
\frac{\partial H_2}{\partial u} \right) R_{\mu} \label{current2}
\end{equation}
where $G$ is an arbitrary function of $\xi, u, u^*$ whereas
$H_1,H_2$ depend only on $u$ and $u^*$. Indeed, such currents are
conserved if the fields satisfy the integrability condition
(\ref{old int cond1}). The importance of this integrable submodel
becomes transparent if one notices that the skyrmions with charges
$Q=\pm1$ as well as the rational map Ansatz, which is widely used
to approximate true numeric solitons
\cite{numerical1}-\cite{numerical4}, belong to it.
\\
On the other hand, it is known that the eikonal equation
constrains the space of solutions in a rather considerable way.
For example, as was mentioned before, there has been found only
one topological soliton in the integrable submodel of the Nicole
model. In the Faddeev-Skyrme-Niemi model the integrable submodel
seems to contain no solitons.
\\
In the case of the Skyrme model the assumption of the eikonal
equation as a constraint results in a restriction of the form of
the $u$ field. Concretely, for the separation of variable ansatz
$\xi =\xi (r)$ and $u=u(\theta ,\phi)$ a $u$ field obeying the eikonal
equation must be a (anti)holomorphic function of the variable
$z=\tan (\frac{\theta}{2}) e^{i\phi}$. If
we combine it with the requirement of the finiteness of the
topological charge, then we find that $u$ must be just a rational
map in $z$ and, as a consequence, skyrmions with higher charges do not
belong to this integrable sector. Because of that, it is
reasonable to seek other integrable submodels, which are defined
by weaker integrability conditions.
%%%%%%%%%%%%%%%%%%%%%%%%%%%%%%%%%%%%%%%%%%%%%%%%
\subsection{New integrable submodel}
%%%%%%%%%%%%%%%%%%%%%%%%%%%%%%%%%%%%%%%%%%%%%%%%
It has been recently demonstrated \cite{weak} that for all
nonlinear theories in Min\-kow\-ski space with two-dimensional
target space there exists an integrability condition which is
weaker than the eikonal equation. Here, we will apply this result
in the context of the Skyrme model.
\\
We begin with the only assumption that
\begin{equation}
\partial_{\mu} \xi \partial^{\mu} u =0 \label{new int cond1}
\end{equation}
whereas, contrary to the standard integrability subsector,
$(\partial u)^2$ is allowed to take arbitrary, in particular
nonzero, values. Thus, we get only the second constraint in
equation (\ref{old int cond1})
\begin{equation}
R_{\mu} \partial^{\mu} u=0. \label{new int cond2}
\end{equation}
The pertinent equations of motion read
\begin{equation}
\partial_{\mu} S^{\mu} = \frac{2}{1+|u|^2} u^* S^{\mu} \partial_{\mu} u
 \label{eq mot u}
\end{equation}
and
\begin{equation}
\partial_{\mu} R^{\mu} = 4 \frac{\sin \xi \cos \xi}{(1+|u|^2)^2}
S^{\mu} \partial_{\mu} u^*. \label{eq mot xi}
\end{equation}
Since the eikonal equation is not imposed,  $S^{\mu}
\partial_{\mu} u \neq 0$ and the left hand side in (\ref{eq mot u})
is not identically zero. One can find that
\begin{equation}
S^{\mu} \partial_{\mu} u = (\partial u)^2(1 -4\lambda^2 (\partial
\xi)^2) \label{rel1}
\end{equation}
The first class of conserved currents can be taken as follows
\begin{equation}
J_{\mu}^{G}= G_u S_{\mu} - G_{u^*} S^*_{\mu}. \label{currents 1}
\end{equation}
In spite of the fact that the form of the currents is identical to
(\ref{current1}) there is, however, a subtle modification. Now,
an arbitrary function $G$ is assumed to depend on the square of the
modulus of $u$,
i.e., $G=G(uu^*, \xi)$.
\\
Then, the divergence reads
$$
\partial^{\mu} J_{\mu}=
G_{uu} \partial^{\mu} u S_{\mu} +G_{uu^*}
\partial^{\mu} u^* S_{\mu} +G_u \partial^{\mu} S_{\mu} $$
\begin{equation}
-G_{u^*u^*} \partial^{\mu} u^* S_{\mu}^* - G_{u^*u}
\partial^{\mu}u S^*_{\mu} - G_{u^*} \partial^{\mu} S^*_{\mu} +G_u
\partial^{\mu} \xi S_{\mu} - G_{u^*} \partial^{\mu} \xi S_{\mu}^*
\label{div j 1}
\end{equation}
Taking into account (\ref{new int cond2}) and observing that
$$S^{\mu}\partial_{\mu} \xi =0, \;\;\; \partial^{\mu} u^* S_{\mu} =
\partial^{\mu} u S^*_{\mu}$$ we get
\begin{equation}
\partial^{\mu} J_{\mu}= G''(u^{*2} \partial^{\mu}u S_{\mu} - u^2
\partial^{\mu}u^*
S^*_{\mu}) +G'(u^* \partial^{\mu} S_{\mu} - u \partial^{\mu}
S^*_{\mu}), \label{div j 2}
\end{equation}
where prime denotes differentiation with respect to the modulus squared.
Using (\ref{eq mot u}) and (\ref{rel1}) one can rewrite it as
$$
\partial^{\mu} J_{\mu}=G''(1-4\lambda^2(\partial \xi)^2)[u^{*2} (\partial
u)^2 - u^2 (\partial
u^*)^2] + $$
\begin{equation}
G' \frac{2}{1+|u|^2} (1-4\lambda^2(\partial \xi)^2) [u^{*2}
(\partial u)^2 - u^2 (\partial u^*)^2]. \label{div j 3}
\end{equation}
Thus, the current is conserved if we assume the weak integrability
condition
\begin{equation}
[u^{*2} (\partial u)^2 - u^2 (\partial u^*)^2]=0. \label{int cond
2}
\end{equation}
Similarly, one can consider the second type of currents, i.e.,
\begin{equation}
J_{\mu}^{(H_1,H_2)}=4\sin \xi \cos \xi (H_1S_{\mu}+H_2S_{\mu}^*)
-(1+|u|^2)^2 \left(\frac{\partial H_1}{\partial u^*} +
\frac{\partial H_2}{\partial u} \right) R_{\mu}. \label{currents
2}
\end{equation}
It can be easily shown that its divergency vanishes if the
functions $H_1$ and $H_2$ are of the form $$H_1=\frac{\partial
h}{\partial u}, \; \; \; H_2=-\frac{\partial h}{\partial u^*},$$
where $h$ is any function of the modulus squared $uu^*$. It strongly
simplifies the currents. Namely, they are reduced to the first
class currents $J^G$.
\\
Let us now summarize the results obtained so far. We have defined
a new integrable sector of the Skyrme model which consists of two
integrability constraints
\begin{equation}
\partial_{\mu} \xi \partial^{\mu} u =0, \; \; \;
u^{*2} (\partial u)^2 - u^2 (\partial u^*)^2=0 \label{new int
cond}
\end{equation}
and two dynamical equations (\ref{eq mot u}), (\ref{eq mot xi}).
The pertinent infinite family of conserved currents is
parameterized by any function $G$, which depends on $uu^*$ and
$\xi$ in  an arbitrary manner
\begin{equation}
J_{\mu}^{G} =   G_u S_{\mu}-G_{u^*} S^*_{\mu} . \label{conserved
current}
\end{equation}
The geometric meaning of the integrability conditions (\ref{new
int cond}) becomes especially well visible if we express the
complex field in terms of the scalars $\Sigma, \Lambda$
\begin{equation}
u=e^{\Sigma +i\Lambda}. \label{decomposition}
\end{equation}
Then, using equations (\ref{new int cond}) we get
\begin{equation}
\partial_{\mu} \Lambda \partial^{\mu} \Sigma=0, \; \; \partial_{\mu}
\Lambda \partial^{\mu} \xi =0, \;
\; \partial_{\mu} \Sigma \partial^{\mu} \xi =0. \label{meaning}
\end{equation}
In other words, in the integrable sector all gradients of the
scalar fields must be mutually perpendicular.
\\
It is worth to notice that such an integrable submodel can be
constructed for a massive generalization of the Skyrme theory
\cite{skyrme_mass} as well. For example, introducing the most
typical massive term
\begin{equation}
V=f^2_{\pi} m^2 \mbox{Tr} (1-U)
\end{equation}
results in a modification of equation (\ref{eq mot xi})
\begin{equation}
\partial_{\mu} R^{\mu} = 4 \frac{\sin \xi \cos \xi}{(1+|u|^2)^2}
S^{\mu} \partial_{\mu} u^* +2m^2 \sin \xi. \label{eq mot xi
massive}
\end{equation}
One can check that it does not influence the definition of the
integrable sector and the conserved currents. This may be of some
interest since the Skyrme model with a massive term serves
considerably better as a model of the low energy QCD, describing
low energy degrees of freedom i.e. hadrons and nucleons with acceptable
accuracy \cite{ANW}, \cite{massive skyrme}.
\\
As was discussed in the previous subsection, all rational maps
(or generally holomorphic functions) obey the standard
integrability condition. Obviously, they satisfy the weaker
condition as well. However, this weaker condition allows for a
much larger class of solutions. For instance, non-holomorphic
functions in the form (\ref{decomposition}) fulfill the
constraint. It may be of some interest since it has been recently
shown that a non-holomorphic Ansatz approximates true Skyrme
solitons considerably better than the standard rational Ansatz
\cite{krusch}, \cite{nonholomic_skyrme}. In particular, the
non-holomorphic approximation of the skyrmion with topological
charge $Q=2$ \cite{krusch} obeys our integrability condition.
%%%%%%%%%%%%%%%%%%%%%%%%%%%%%%%%%%%%%%%%%%%%%%%%
\subsection{Integrable chiral model}
%%%%%%%%%%%%%%%%%%%%%%%%%%%%%%%%%%%%%%%%%%%%%%%%
One might ask whether it is possible to further relax the
integrability conditions (\ref{new int cond}). In fact, one can
construct a chiral $SU(2)$ model which is integrable even if the
second constraint in (\ref{new int cond}) is neglected. In other
words, there are no additional requirements for the the complex
field $u$. This property, as it will be shown below, allows for
the existence of solitons with an arbitrary value of the
topological charge.
\\
The model we are going to focus on is given by the following
Lagrangian density
\begin{equation}
L=-f(\xi) g(u,u^*) \left[ K_{\mu}
\partial^{\mu} u^*\right]^{\frac{3}{4}}+\left[(\partial_{\mu}
\xi)^2 \right]^{\frac{3}{2}}, \label{model toy}
\end{equation}
where $f$ and $g$ are arbitrary, differentiable functions
depending on $\xi$ and $u,u^*$ respectively. Additionally
\begin{equation}
K_{\mu} = (\partial_{\nu} u \partial^{\nu} u^*) \partial_{\mu} u -
(\partial_{\nu} u \partial^{\nu} u) \partial_{\mu} u^*. \label{def
K_new}
\end{equation}
The fractional exponent presented in the upper formula is
understood as follows
\begin{equation}
\left[ K_{\mu}
\partial^{\mu} u^*\right]^{\frac{3}{4}} \equiv  K_{\mu}
\partial^{\mu} u^* | K_{\mu}
\partial^{\mu} u^* |^{-\frac{1}{4}}. \label{def fractional}
\end{equation}
It can be observed that such a form of the model guarantees also
an interesting way of circumventing Derrick's theorem since the
energy is invariant under scaling transformations. This attempt
was originally proposed 30 year ago by Deser et al \cite{deser} to
consider the theory of pion fields $\pi^i$ and further developed
by many authors (see Nicole model \cite{nicole},
Aratyn-Ferreira-Zimerman model \cite{aratyn2} and their
generalizations \cite{afz gen1}, \cite{afz gen2}). Let us notice
that our model can be treated as the modified AFZ model coupled in
a non-minimal way (via the dielectric function $f$) with a
non-standard scalar field \footnote{ Models with a non-canonical
kinetic term for the scalar field have been recently widely
discussed in the cosmology. They have been applied in the context
of inflation \cite{inflation}, dark matter \cite{dark} as well as
the modified Newtonian dynamics (MOND) \cite{mond}. For the
detailed studies of the kinetic term occurring in our model
(\ref{model toy}) see \cite{kinetic}.}.
\\
In principle, one may rewrite this model in terms of the original
chiral field $U$ but in practice such a Lagrangian would take a
completely illegible form.
\\
The pertinent equation of motion for the complex scalar field
reads
\begin{equation}
\frac{3}{2} \partial_{\mu} \left[ f g \left[ K_{\mu}
\partial^{\mu} u^*\right]^{-\frac{1}{4}} K^{\mu} \right] -g_{u^*}
f \left[ K_{\mu}
\partial^{\mu} u^*\right]^{\frac{3}{4}}=0 \label{eq u1}
\end{equation}
where $g_{u^*} \equiv \partial_{u^*} g$.
This equation can be simplified to the following expression
\begin{equation}
\partial_{\mu} \left[ f g^{\frac{1}{3}} \left[ K_{\mu}
\partial^{\mu} u^*\right]^{-\frac{1}{4}} K^{\mu} \right] =0 \label{eq
u2}
\end{equation}
or in the most compact form
\begin{equation}
\partial_{\mu} \mathcal{K}^{\mu} =0, \label{eq u3}
\end{equation}
where
\begin{equation}
\mathcal{K}^{\mu}=f g^{\frac{1}{3}} \left[ K_{\mu}
\partial^{\mu} u^*\right]^{-\frac{1}{4}} K^{\mu}. \label{def math
K}
\end{equation}
The second independent equation of motion, for the real scalar
field $\xi$, is
\begin{equation}
3\partial_{\mu} \left[ \left[(\partial_{\mu} \xi)^2
\right]^{\frac{1}{2}} \partial^{\mu} \xi \right] +f_{\xi} g
\left[ K_{\mu}
\partial^{\mu} u^*\right]^{\frac{3}{4}} =0. \label{eq xi1}
\end{equation}
In order to establish the integrability property of the model we
introduce the following currents
\begin{equation}
J_{\mu} = \mathcal{K}_{\mu} G_u - \mathcal{K}^*_{\mu} G_{u^*},
\label{curr integ}
\end{equation}
where $G$ is an arbitrary function of $u, u^*$ and $\xi$. Taking
into account the equations of motion it can be proved that these
currents are conserved if we assume only one constraint (\ref{new
int cond1}). It is due to the fact that the expression
$\mathcal{K}_{\mu}
\partial^{\mu}u=0$ is a mathematical identity which does not
restrict the form of the complex field.
\\
In order to find topologically nontrivial solutions of these
dynamical equations we will consider only time-independent
configurations and assume the Ansatz for the fields
\begin{equation}
u=u(\phi, \theta)= e^{in\phi} h (\theta) \label{anzatz u}
\end{equation}
and
\begin{equation}
\xi =\xi (r), \label{anzatz xi}
\end{equation}
where the spherical coordinates $(r,\theta,\phi)$ have been
introduced. Here $n$ is an integer number. Of course, such an
Ansatz obeys the integrability condition. Thus, the obtained
solitons will belong to the integrable submodel.
\\
One can easily find that
\begin{equation}
\vec{K} \cdot \nabla u^*=4 \left( \frac{nhh_{\theta}}{\sin \theta
r^2} \right)^2 \label{vecK u}
\end{equation}
and as a consequence
\begin{equation}
\vec{\mathcal{K}}=\sqrt{2} g^{1/3} f \left(
\frac{nhh_{\theta}}{\sin \theta}\right)^{\frac{1}{2}}
\frac{e^{in\phi}}{r^{2}} \left[ 0, \frac{nh}{\sin \theta},
ih_{\theta} \right]. \label{math cal1}
\end{equation}
Substituting these formulas into the pertinent field equation we
get the following second order, ordinary differential equation for
the function $h$
\begin{equation}
\partial_{\theta} \left[ nh g^{1/3} \left( \frac{nhh_{\theta}}{\sin \theta}
\right)^{\frac{1}{2}} \right] -nh_{\theta}g^{1/3}(h) \left(
\frac{nhh_{\theta}}{\sin \theta} \right)^{\frac{1}{2}} =0.
\label{eq h1}
\end{equation}
It can be simplified to this expression
\begin{equation}
\partial_\theta \left[ g^{1/3} \left( \frac{hh_{\theta}}{\sin \theta}
\right)^{\frac{1}{2}} \right] =0, \label{eq h2}
\end{equation}
which possesses the obvious solution
\begin{equation}
g^{1/3} \left( \frac{hh_{\theta}}{\sin \theta}
\right)^{\frac{1}{2}} = \mu, \label{eq h3}
\end{equation}
where $\mu$ is a positive constant.
\\
Taking into account the Ansatz and the upper obtained solution
(\ref{eq h3}) we are able to rewrite the equation of motion for
$\xi$ in the form
\begin{equation}
\frac{3}{r^2} \partial_r \left[ r^2 |\xi_r| \xi_r \right]
-f_{\xi}  \frac{(2n)^{3/2} \mu^{3}}{r^3}=0. \label{eq xi2}
\end{equation}
After introducing a new variable
\begin{equation}
x=\ln r \label{zamian x}
\end{equation}
we derive the following equation
\begin{equation}
\partial_x \left[ |\xi_x| \xi_x \right] -f_{\xi} \frac{
(2n)^{3/2} \mu^{3}}{3}=0. \label{eq xi3}
\end{equation}
Fortunately it can be integrated for any function $f$. The
solution is
\begin{equation}
\left( \xi_x \right)^3 = \frac{(2n)^{3/2}}{2} \mu^{3} f(\xi).
\label{eq xi4}
\end{equation}
Both formulas (\ref{eq h3}) as well as (\ref{eq xi4}) can be
integrated, at least using standard numerical methods, for all
reasonable functions $g$ and $f$ leading to the following general
solutions
\begin{equation}
\int g^{2/3} h dh = -\mu^2 \cos \theta + \alpha_0 \label{sol gen
h}
\end{equation}
and
\begin{equation}
\int \frac{d\xi}{\sqrt[3]{f}} = \frac{(2n)^{1/2}}{\sqrt[3]{2}} \mu
\ln \frac{r}{r_0}, \label{sol gen xi}
\end{equation}
where $\alpha_0, r_0$ are integration constants.
\\
Let us now analyze in detail a particular case with the following
forms of the coupling functions
\begin{equation}
g(u,u^*) = \frac{1}{(1+|u|^2)^3} \label{form g}
\end{equation}
and
\begin{equation}
f(\xi) = \sin^3 \xi. \label{form xi}
\end{equation}
The corresponding Lagrangian density can be expressed as follows
\begin{equation}
L=-\sin^3 \xi \; [\vec{n} \cdot (\partial_{\mu} \vec{n} \times
\partial_{\nu} \vec{n} )]^{\frac{3}{4}} + [(\partial_{\mu}
\xi)^2]^{\frac{3}{2}}, \label{exact form lagr}
\end{equation}
where $\vec{n}=\vec{\xi}/\xi$ is a three component unit vector.
\\
Then
\begin{equation}
\frac{hh_{\theta}}{(1+h^2)^2}=\mu^2 \sin \theta . \label{eq h4}
\end{equation}
Hence, the solution is given as follows
\begin{equation}
h^2= \frac{ (1+\alpha_0+\mu^2) \sin^2 \frac{\theta}{2} +
(1+\alpha_0-\mu^2) \cos^2 \frac{\theta}{2} }{(-\alpha_0-\mu^2)
\sin^2 \frac{\theta}{2} +(-\alpha_0+\mu^2)\cos^2
\frac{\theta}{2}}. \label{sol h1}
\end{equation}
This solution reduces to a well-known form if the parameters
satisfied
\begin{equation}
1+\alpha_0 -\mu^2=0 \; \; \; \mu^2 +\alpha_0=0. \label{special mu}
\end{equation}
That is
\begin{equation}
\alpha_0= -\frac{1}{2} \; \; \; \mu^2= \frac{1}{2}. \label{special
mu2}
\end{equation}
Indeed, then we get
\begin{equation}
h = \tan \frac{\theta}{2}. \label{sol h2}
\end{equation}
In other words we obtained the following complex scalar field
\begin{equation}
u(\theta, \phi) = e^{in\phi} \tan \frac{\theta}{2}. \label{sol u}
\end{equation}
One can observe that if we take into account the stereographic
projection then we find
\begin{equation}
\vec{n} = \left( \cos n\phi \sin \theta, \sin n\phi \sin \theta,
\cos \theta \right). \label{sol vec xi}
\end{equation}
The remaining field equation
\begin{equation}
\xi_x= \frac{\sqrt{n}}{\sqrt[3]{2}} \sin \xi \label{eq xi5}
\end{equation}
can be solved as well, providing the solution
\begin{equation}
\xi = 2 \arctan \left( \frac{r}{r_0} \right)^{\frac{\sqrt{n}
}{\sqrt[3]{2}}}. \label{sol xi}
\end{equation}
The topological charge of the obtained configuration can be
computed from the standard expression
\begin{equation}
Q=\frac{1}{4\pi^2} \int 2i \frac{du \wedge du^* \wedge
d\xi}{(1+|u|^2)^2}. \label{top charg1}
\end{equation}
Inserting our solutions (\ref{sol u}), (\ref{sol xi}) into
(\ref{top charg1}) one gets that
\begin{equation}
Q=n \int \frac{d \xi}{\pi} = n \frac{\xi(0)-\xi(\infty)}{\pi} =
-n.
\end{equation}
Therefore the solution represents a chiral field with arbitrary
value of the baryon number.
\\
Let us now compute the corresponding total energy. In the case of
static configurations it is given by the formula
\begin{equation}
E=\int \; \left( f(\xi) g(u,u^*) \left[ \vec{K} \nabla u^*
\right]^{\frac{3}{4}} + \left[ (\nabla \xi)^2
\right]^{\frac{3}{2}} \right) \;  d^3 x. \label{static energy}
\end{equation}
In particular, for the example considered above, it can be
rewritten as
\begin{equation}
E= \int  \left[ \left( \frac{n}{2} \right)^{\frac{3}{2}}
\frac{sin^3 \xi}{r^3} + \left( \xi'_r \right)^3 \right] r^2 \sin
\theta d\phi d\theta dr. \label{energy1}
\end{equation}
Taking into account equation (\ref{eq h3}) we get
\begin{equation}
E=4\pi \cdot 3 \int_0^{\infty} dr r^2 \xi^3_r. \label{energy2}
\end{equation}
Finally, using solution (\ref{sol xi}) one finds that
\begin{equation}
E=\frac{3\pi^2}{\sqrt[3]{2}} \; n. \label{energy total}
\end{equation}
As one could expect the energy of the obtained configurations is
proportional to the topological charge.
\\
{\it Remark:} It is easy to observe that solutions of the model
(\ref{model toy}) with other coupling function $g(u,u^*)$ have the
same topological properties. Of course, it is true only if the
pertinent boundary conditions for $h$ and $\xi$ are fulfilled.
Namely,
\begin{equation}
\xi \longrightarrow \left\{
\begin{array}{cc}
0 \mbox{ or } \pi & \mbox{ if } r \rightarrow 0 \\
\pi \mbox{ or } 0 & \mbox{ if } r \rightarrow \infty
\end{array}
\right. \; \; \; \; h \longrightarrow \left\{
\begin{array}{cc}
0 \mbox{ or } \infty & \mbox{ if } \theta \rightarrow 0\\
\infty \mbox{ or } 0 & \mbox{ if } \theta \rightarrow \infty \; .
\end{array}
\right. \label{cond boundary}
\end{equation}
For instance, if one considers the function $g$ in the following
form
\begin{equation}
g(u,u^*)=\frac{1}{(1+|u|^2)^p}, \label{other g}
\end{equation}
where $p > 3/2$. Then the general solution reads
\begin{equation}
h^2= \left( \frac{3}{2(2p-3)\left( \mu^2 cos \theta -\alpha_0
\right)} \right)^{\frac{3}{2p-3}} -1. \label{sol other}
\end{equation}
The topologically interesting configuration is obtained if the
constants
\begin{equation}
\mu^2=\frac{3\cdot 2^{\frac{3-2p}{3}} }{4(2p-3)}, \;\;\;
\alpha_0=-\mu^2. \label{other const}
\end{equation}
Indeed,
\begin{equation}
h^2=\left( \frac{2}{\cos \theta +1} \right)^{\frac{3}{2p-3}}-1
\label{other top sol}
\end{equation}
is a function which satisfies the required boundary conditions
(\ref{cond boundary}). The energy-charge relation remains
unchanged. Assuming other forms of the function $g$ it is possible
to construct more complicated but still topologically nontrivial
solutions.
\\
{\it Remark:} In order to derive solutions with fractional
topological charge one can consider the following coupling
function
\begin{equation}
f (\xi)=\sin^3 \left(\frac{\xi}{q} \right), \label{fract coupl}
\end{equation}
where $q$ is a positive parameter.  The function $g$ is assumed in
the standard form (\ref{form g}) providing the previously
described complex field (\ref{sol u}). However, the real field
$\xi$ is not longer a map onto the segment $[0,\pi]$. Indeed, now
it is given by the expression
\begin{equation}
\xi = 2q \arctan \left( \frac{r}{r_0}\right)^{
\frac{\sqrt{n}}{q\sqrt[3]{2}} } \label{frac sol xi}
\end{equation}
i.e., $\xi \in[0,q\pi]$. Therefore, the original chiral field $U$
does not cover the whole target space providing an arbitrary, in
general noninteger, value of the pertinent topological index
\begin{equation}
Q=-nq. \label{frac top charg}
\end{equation}
Nonetheless, the energy is still finite
\begin{equation}
E=3 \sqrt[3]{2}\pi^2 nq. \label{frac energy}
\end{equation}
As one sees, the energy is proportional to $Q$ even if it takes
noninteger value.
\\
The existence of finite energy solutions with any noninteger
value of the topological index can indicate the the obtained
chiral solitons are unstable. This is in accordance with Ref.
\cite{kinetic} where we have shown that the static Hopf solitons
in the Aratyn-Ferreira-Zimerman model have analogous stability problems.
%%%%%%%%%%%%%%%%%%%%%%%%%%%%%%%%%%%%%%%%%%%%%%%%%%%%%%%%%%%%%%
\section{$CP^n$ model}
%%%%%%%%%%%%%%%%%%%%%%%%%%%%%%%%%%%%%%%%%%%%%%%%%%%%%%%%%%%%%%
%%%%%%%%%%%%%%%%%%%%%%%%%%%%%%%%%%%%%%%%%%%%%%%%%%%%%%%%%%%%%%
\subsection{Standard integrable submodel}
%%%%%%%%%%%%%%%%%%%%%%%%%%%%%%%%%%%%%%%%%%%%%%%%%%%%%%%%%%%%%%
Let us now investigate the $CP^n$ model in an arbitrary
dimensional space-time \cite{integrability2}, \cite{fujii cpn},
\cite{fujii gras1}, \cite{fujii gras2}
\begin{equation}
L=\frac{(1+u^{\dag} \cdot u)(\partial_{\mu} u^{\dag} \cdot
\partial^{\mu} u) - (u^{\dag} \cdot \partial_{\mu} u
)(\partial^{\mu} u^{\dag} \cdot u)}{(1+u^{\dag} \cdot u)^2},
\label{cpn model}
\end{equation}
where $u$ is a column of $n$ complex fields and $u^{\dag} \cdot u
\equiv u^*_au^a$, $a=1..n$. The pertinent equations of motions
read
\begin{equation}
(1+u^{\dag} \cdot u) \partial^2 u_a = 2 (u^{\dag} \cdot
\partial_{\mu} u ) \partial^{\mu} u_a. \label{cpn eqmot1}
\end{equation}
As it was established in \cite{integrability2} this model also
possesses an integrable sector, given by the dynamical equations
\begin{equation}
\partial^2 u_a=0, \;\; a=1..n   \label{cpn integr sector1}
\end{equation}
and the set of integrability conditions
\begin{equation}
\partial_{\mu} u_a
\partial^{\mu} u_b=0, \;\; 1 \leq a \leq b \leq n. \label{cpn integr sector2}
\end{equation}
In fact, if we assume (\ref{cpn integr sector1}), (\ref{cpn integr
sector2}) the following currents are conserved
\begin{equation}
J_{\mu}= \frac{\partial \mathcal{F}}{\partial u_i}
\partial_{\mu} u_i -\frac{\partial \mathcal{F}}{\partial u_i^*}\partial_{\mu}
u_i^*, \label{cpn current}
\end{equation}
where $\mathcal{F}$ is an arbitrary function of $u_1,u_1^*...u_n,
u_n^*$. Applying the decomposition of the complex field used
before
\begin{equation}
u_a=e^{\Sigma_a+i\Lambda_a}, \;\;\; a=1...n \label{decomp cpn}
\end{equation}
we can find a geometric interpretation of the integrability
conditions
\begin{equation}
(\partial_{\mu} \Lambda_a )^2 = (\partial_{\mu} \Sigma_b )^2,
\end{equation}
\begin{equation}
\partial_{\mu} \Lambda_a \partial^{\mu} \Sigma_b =0, \;\; 1 \leq a, b \leq
n.
\end{equation}
Interestingly enough, in $(2+1)$ dimensional space-time, the
standard integrable submodel constitutes the BPS sector of the
$CP^n$ model. That is, the solutions of the integrable subsystem
(\ref{cpn integr sector1}), (\ref{cpn integr sector2}) are nothing
else but the very well-known BPS solutions of the $CP^n$ model
\cite{cpn_sol1}-\cite{cpn_sol_time}
\begin{equation}
U=\frac{f}{|f|}, \label{zakrz sol gen}
\end{equation}
where $f$ is a vector of $f_a=f_a(z)$ (anti)holomorphic functions
of $z=x+iy$. Indeed, due to the conformal symmetry the general
solution of the integrable submodel is just a collection of
(anti)holomorphic functions
\begin{equation}
u^a=u^a(z) \;\; \mbox{or} \;\; u^a=u^a(\bar{z}). \label{2dim sol}
\end{equation}
Then, using the local $U(1)$ invariance one can express the BPS
solutions of the $CP^n$ model in the following form
\begin{equation}
U=\frac{1}{\sqrt{1+u^{a*}u^a}} \left(
\begin{array}{c}
 1 \\ u
\end{array} \right). \label{zakrz sol}
\end{equation}
Of course, as one usually wants to deal with objects carrying
finite topological charge the holomorphic function must be
restricted to arbitrary rational functions.
\\
In more than two spatial dimensions the conformal symmetry does
not occur and the chance for the existence of an integrable
submodel with a nontrivial set of solutions becomes limited. It is
easy to see that a general solution of the integrability
conditions is $u_a=A_a f(u)$, where $A_a$ are complex constants
and $f$ is any function of $u$, which satisfies the $(d+1)$
dimensional scalar eikonal equation
$$
\partial_{\mu} u
\partial^{\mu} u^*=0.$$ If one combines this result with the fact
that such configurations should obey the dynamical equation
(\ref{cpn integr sector1}) as well, then one may expect that the
chances for nontrivial solutions of the integrable sector are
highly reduced. It, of course, restricts in a quite considerable
way the range of applications of the obtained integrable sectors.
%%%%%%%%%%%%%%%%%%%%%%%%%%%%%%%%%%%%%%%%%%%%%%%%%%%%%%%%%%%%%%
\subsection{New integrable submodels}
%%%%%%%%%%%%%%%%%%%%%%%%%%%%%%%%%%%%%%%%%%%%%%%%%%%%%%%%%%%%%%
In this section we construct new integrable sectors in the $CP^n$
model. In general, such alternative submodels are found if much
less restrictive constraints are imposed.
\\
We consider currents  similar to (\ref{cpn current})
\begin{equation}
J_{\mu}^{(i)}= \frac{\partial \mathcal{F}}{\partial u_i}
\partial_{\mu} u_i -\frac{\partial \mathcal{F}}{\partial u_i^*}\partial_{\mu}
u_i^*, \;\;\; i=1...n \label{cpn current new1}
\end{equation}
but the function $\mathcal{F}$ is assumed to depend on the field
variables in a specific manner. Namely,
\begin{equation}
\mathcal{F}=\mathcal{F}
(u^*_1u_1+...+u^*_ku_k,u^*_{k+1}u_{k+1},...,u^*_nu_n). \label{F
form1}
\end{equation}
Of course, we expect that the currents are conserved i.e.
\begin{equation}
\partial^{\mu} J_{\mu}^{(i)}=0, \;\;\; i=1...n.
\end{equation}
Thus, taking into account the field equations (\ref{cpn eqmot1}),
we get these integrability conditions
\begin{equation}
\sum_{b=1}^k (u^*_au^*_b \partial_{\mu} u_a \partial^{\mu} u_b -
u_au_b \partial_{\mu} u_a^* \partial^{\mu} u_b^* ) =0, \;\;\;
a=1...n
\end{equation}
\begin{equation}
\sum_{b=1}^k (u^*_au_b \partial_{\mu} u_a \partial^{\mu} u_b^* -
u_au_b^* \partial_{\mu} u_a^* \partial^{\mu} u_b ) =0, \;\;\;
a=1...n
\end{equation}
\begin{equation}
u^*_au^*_b \partial_{\mu} u_a \partial^{\mu} u_b - u_au_b
\partial_{\mu} u_a^* \partial^{\mu} u_b^*=0, \;\;\; b=k+1,...,n
\;\;\; a=1...n
\end{equation}
\begin{equation}
u^*_au_b \partial_{\mu} u_a \partial^{\mu} u_b^* - u_au_b^*
\partial_{\mu} u_a^* \partial^{\mu} u_b=0, \;\;\; b=k+1,...,n
\;\;\; a=1...n.
\end{equation}
Once again, it is convenient to rewrite them using (\ref{decomp
cpn}). Then,
\begin{equation}
\partial_{\mu} \Lambda_a \left( \sum_{b=1}^k \partial^{\mu}
\Sigma_b \right)=0, \;\;\; \partial_{\mu} \Sigma_a \left(
\sum_{b=1}^k
\partial^{\mu} \Lambda_b \right)=0, \;\;\; a=1...k
\end{equation}
\begin{equation}
\partial_{\mu} \Lambda_a \partial^{\mu}
\Sigma_b =0, \;\;\; \partial_{\mu} \Sigma_a \partial^{\mu}
\Lambda_b=0, \;\;\; a=1...n, b=k+1,...,n.
\end{equation}
As it was described in the Skyrme model in the previous section
(or in the Faddeev-Niemi model in \cite{weak}),
in an integrable sector gradients of the pertinent real scalar
fields (real coordinated on the target space) must be
perpendicular. However, it does not mean that all gradients are
mutually perpendicular. Varying the form of the $\mathcal{F}$
function we get that for some of the gradients it is sufficient to
be perpendicular to a sum of others.
\\
To obtain even weaker integrability conditions, and in a
consequence a submodel with a larger set of solutions, one can
take under consideration the following current
\begin{equation} \label{CPn-sum-cur}
J_{\mu} = \sum_{i=1}^n  \left( \frac{\partial
\mathcal{F}}{\partial u_i}
\partial_{\mu} u_i -\frac{\partial \mathcal{F}}{\partial u_i^*}\partial_{\mu}
u_i^* \right).
\end{equation}
In this case the integrability conditions read
\begin{equation}
\left( \sum_{a=1}^k \partial_{\mu} \Lambda_a \right) \left(
\sum_{b=1}^k
\partial^{\mu} \Sigma_b \right)=0,
\end{equation}
\begin{equation}
\partial_{\mu} \Lambda_a \left( \sum_{b=1}^k \partial^{\mu} \Sigma_b
\right)+ \partial_{\mu} \Sigma_a \left( \sum_{b=1}^k
\partial^{\mu} \Lambda_b \right)=0, \;\;\; a=k+1,...,n
\end{equation}
\begin{equation}
\partial_{\mu} \Lambda_a  \partial^{\mu} \Sigma_b
+ \partial_{\mu} \Sigma_a
\partial^{\mu} \Lambda_b =0, \;\;\; a=k+1,...,n \;\; b=k+1,...,n.
\end{equation}
To conclude, there is a family of integrable sectors of $CP^n$
model. Integrability conditions vary from the strongest ones (the
most restrictive)
\begin{equation}
\partial_{\mu} \Lambda_a  \partial^{\mu} \Sigma_b=0, \;\;
a,b=1...n, \label{strongest}
\end{equation}
when $\mathcal{F}=\mathcal{F}(u_1u^*_1,...,u_nu_n^*)$ to the
weakest ones (the lest restrictive)
\begin{equation}
\left( \sum_{a=1}^n \partial_{\mu} \Lambda_a \right) \left(
\sum_{b=1}^n
\partial^{\mu} \Sigma_b \right)=0,
\end{equation}
if $\mathcal{F}=\mathcal{F}(u_1u^*_1+...+u_nu_n^*)$ and take all
possible intermediate cases. Obviously, the full integrable
submodel consists of the field equations (\ref{cpn eqmot1}) and a
choice of an integrability condition, as discussed above.
\\
It should be underlined that the construction of such integrable
sectors of the $CP^n$ model remains unchanged if we add a
potential term to the Lagrangian \cite{cpn_pot}.
\\ \\
Let us again interpret the derived integrable sectors in the
context of $CP^n$ model in $(2+1)$ dimensions. One can check that
the most restrictive constraint (\ref{strongest}) possesses, in
addition to standard ones (\ref{2dim sol}), the following solution
\begin{equation}
\Lambda_a=\Lambda_a \left(\frac{z}{z^*} \right), \;\;\;
\Sigma_a=\Sigma_a(zz^*), \label{strongest sol}
\end{equation}
where $\Lambda_a, \Sigma_a$ are arbitrary functions depending on
phase and modulus respectively. On the other hand it is known that
the $CP^n$ model has also non-BPS solutions, which can be
constructed from the BPS solitons by acting with a projective
operator \cite{cpn_sol1}-\cite{cpn_sol_time}
\begin{equation}
U=\frac{P^k_+f}{|P^k_+f|}, \label{zakrz nonBPS}
\end{equation}
where
$$P_+f=\partial_z f -
(f^{\dagger}\partial_zf)\frac{f}{|f|^2}, \;\;\; k=0,..,n.$$
These
non-BPS solutions (or at least some of them) obey the integrability
conditions. As an example let us consider $f=(1,z,z^2)$, then for
$k=1$
\begin{equation}
U=\frac{1}{\sqrt{1+4|z|^2+6|z|^4+5|z|^6+|z|^8}} \left(
\begin{array}{c}
z^*(1+2|z|^2)  \\
|z|^4-1 \\
-z (2+|z|^2)
\end{array} \right)_. \label{nonBPS eg}
\end{equation}
Now, after expressing it by means of the parameterization
(\ref{zakrz sol}) we get that the pertinent scalar complex
functions are in the form (\ref{strongest sol}). Since all non-BPS
solutions by construction fulfill the dynamical equation of
motions (\ref{cpn eqmot1}), we can conclude that new integrable
sectors of the $CP^n$ model in two spatial dimensions consist of
BPS as well as non-BPS states.
\\
As we see, in $(2+1)$ dimensions, there is a striking
correspondence between integrability and BPS property of
solutions. The strong integrable sector is just the BSP sector
whereas weaker integrable sectors contain also non BPS solutions.
In higher dimensions we usually do not have any BPS solitons,
however, the strong as well as weaker integrable sectors are still
well-defined. Thus, integrability is an important tool in
investigating nonlinear models in higher dimensions.
\\ \\
An algebraic meaning of the family of integrable sector can be
easily found if one considers volume preserving diffeomorphisms of
the volume 2n-form on the $CP^n$ target space
\begin{equation}
\Omega=g(uu^*_1,...,u_nu^*_n)du_1\wedge du^*_1\wedge...\wedge d
u_n\wedge d u^*_n. \label{2n form}
\end{equation}
In other words we are interested in transformations $u_a
\rightarrow v_a(u_1,u^*_1,...,u_n,u^*_n)$ which leave the 2n-from
unchanged
$$
\Omega=g(uu^*_1,...,u_nu^*_n)du_1\wedge du^*_1\wedge...\wedge d
u_n\wedge du^*_n= $$
\begin{equation}
g(vv^*_1,...,v_nv^*_n)du_1\wedge dv^*_1\wedge...\wedge d v_n\wedge
dv^*_n. \label{2n form inv}
\end{equation}
Let us discuss infinitesimal transformation $u_a \rightarrow u_a
+\epsilon_a$. Then
\begin{equation}
du_a \rightarrow du_a+\epsilon^{a}_{u_b}du_b
+\epsilon^{a}_{u_b^*}du_b^*, \label{inf trans diff}
\end{equation}
where summation over indices $b$ is performed. Now, the invariance
condition (\ref{2n form inv}) gives
$$
g(\epsilon^1_{u_1}+\epsilon^{*1}_{u^*_1}+...+\epsilon^n_{u_n}+
\epsilon^{*n}_{u^*
_n}) +$$
\begin{equation}
(\partial_1 g) (u_1\epsilon^{*1} +u^*_1\epsilon^1) + ...+
(\partial_n g)(u_n\epsilon^{*n} +u^*_n\epsilon^n)=0. \label{inv cond1}
\end{equation}
Here $\epsilon^a_{u_a} \equiv \partial_{u_a} \epsilon^{a}$ and
$\partial_1 g\equiv \partial_{u_1 u^*_1} g$, etc. After introducing
\begin{equation}
\epsilon^a=g^{-1}F_{u^*_a} \label{inv cond def}
\end{equation}
we can rewrite equation (\ref{inv cond1}) in a very compact form
\begin{equation}
\left(
\partial_{u_1}\partial_{u^*_1}+...+\partial_{u_n}\partial_{u^*_n}
\right) (F+F^*)=0. \label{inv cog_{u^*}nd2}
\end{equation}
The general solution reads
\begin{equation}
F+F^*=\sum_i \zeta^i (u_1^{\sigma_i},...,u_n^{\sigma_i}),
\label{inv cond sol1}
\end{equation}
where $\sigma_i=1,2$ and $u^1 \equiv u $, $u^2 \equiv u^*$.
However, from our point of view it is sufficient to investigate
only pure imaginary solutions
\begin{equation}
F+F^*=0. \label{inv cond sol2}
\end{equation}
Thus, $F=iG$ where $G=G(u_1,u^*_1,...,u_n,u^*_n)$ is an arbitrary
real function. It follows that all volume preserving
diffeomorphisms are generated by the vector fields
\begin{equation}
v^G=ig^{-1} \sum_{a=1}^n \left( G_{u^*_a} \partial_{u_a} - G_{u_a}
\partial_{u^*_a} \right), \label{inv cond vector}
\end{equation}
which satisfy the Lie algebra
\begin{equation}
[v^{G_1},v^{G_2}]=v^{G_3}, \;\; G_3=\sum_{a=1}^n \left(
G_{1u^*_a}G_{2u_a} -G_{1u_a}G_{2u^*_a} \right). \label{lie alg}
\end{equation}
In order to find Abelian subalgebras we assume that $G_{1,2}$ take
the following form
\begin{equation}
G=G(u_1u^*_1+...+u^ku^*_k, u_{k+1}u^*_{k+1},...,u_nu^*_n),
\label{fun abelian lie}
\end{equation}
where $k=1..n$ is a fixed number. Indeed, the corresponding
commutator vanishes. Thus for each value of $k$ we get an Abelian
subalgebra $\mathcal{H}_k$. It is straightforward to see that
these subalgebras form a tower of algebras
\begin{equation}
\mathcal{H}_1 \supset \mathcal{H}_{2} \supset ...\supset
\mathcal{H}_{n-1}. \label{tower}
\end{equation}
Further, the vector fields $v^G$ can be identified with the Noether
charges of the currents (\ref{CPn-sum-cur}), once the identification
$G\to {\cal F}$ is made.
%%%%%%%%%%%%%%%%%%%%%%%%%%%%%%%%%%%%%%%%%%%%%%%%%%%%%%%%%%%%%%
\section{Conclusions}
%%%%%%%%%%%%%%%%%%%%%%%%%%%%%%%%%%%%%%%%%%%%%%%%%%%%%%%%%%%%%%
We have shown that the method, developed in \cite{weak} and
originally devoted to the construction of integrable submodels for
$S^2$ valued field theories, can be easily adapted for models with
different target space, and defined in an arbitrary
dimensional space-time.
\\
The interesting point is that the new derived integrable sectors
of Skyrme as well as $CP^n$ models are defined by integrable
conditions which are considerably weaker than the standard ones.
Thus, one might expect that the set of solutions of the integrable
subsystems is larger. In particular, it may cure the problem of
the apparent non-existence of soliton solutions with a topological
number largen than one in the standard integrable sector of the
Skyrme model. However, the problem whether the new integrable
submodel possesses such topological solutions is still an open
question.
\\
Moreover, a correspondence between BPS sector and strong
integrability in the $(2+1)$ dimensional $CP^n$ model
has been established. In addition, we have shown that weak
integrable sectors support non-BPS solutions. Such a relation
holds only in three dimensional space-time but it may indicate
that in higher dimensions integrable submodels can play the role
of BPS sectors. Therefore, one can hope that the construction of
integrable submodels might compensate for the non-existence of BPS
sectors and provide a useful tool for constructing solitons.
%%%%%%%%%%%%%%%%%%%%%%%%%%%%%%%%%%%%%%%%%%%%%%%%%%%%%%%
\section{Acknowledgements}
%%%%%%%%%%%%%%%%%%%%%%%%%%%%%%%%%%%%%%%%%%%%%%%%%%%%%%%
C. A. and J.S.-G. thank MCyT (Spain) and FEDER (FPA2005-01963),
and Incentivos from Xunta de Galicia. A.W. gratefully acknowledges
Departamento de Fisica de Particulas, Universidad de Santiago for
hospitality. Further, C. A. acknowledges support from the Austrian
START award project FWF-Y-137-TEC and from the FWF project P161 05
NO 5 of N.J. Mauser.

\end{document}